\documentclass[english,ngerman]{bvm} 

\bvmdef\articlenumber{0000}

\bvmdef\type{V}

\date{}

\usepackage[font=small,labelfont=bf,skip=5pt]{caption}

\usepackage{fancyhdr}
\fancypagestyle{specialfooter}{%
  \fancyhf{}
  
  \fancyfoot[L]{\scriptsize{$\copyright$ Bildverarbeitung f\"ur die Medizin Workshop 2022 (BVM 2022), scheduled for 20-22 March 2022 in Heidelberg, Germany.}}
}
\title{A Keypoint Detection and Description Network Based on the Vessel Structure for Multi-Modal Retinal Image Registration}

\titlerunning{Multi-Modal Retinal Image Registration}

\author{Aline Sindel$^1$, Bettina Hohberger$^2$, Sebastian Fassihi Dehcordi$^2$, Christian~Mardin$^2$, Robert L{\"a}mmer$^2$, Andreas Maier$^1$, Vincent Christlein$^1$}

\authorrunning{Sindel et al.}

\institute{$^1$Pattern Recognition Lab, FAU Erlangen-N\"urnberg\\ $^2$Department of Ophthalmology, Universit\"atsklinikum Erlangen}

\email{aline.sindel@fau.de}

\begin{document}

%
\selectlanguage{english}

\maketitle
\thispagestyle{specialfooter}

\begin{abstract}
Ophthalmological imaging utilizes different imaging systems, such as color fundus, infrared, fluorescein angiography, optical coherence tomography (OCT) or OCT angiography.
Multiple images with different modalities or acquisition times are often analyzed for the diagnosis of retinal diseases.
Automatically aligning the vessel structures in the images by means of multi-modal registration can support the ophthalmologists in their work. Our method uses a convolutional neural network to extract features of the vessel structure in multi-modal retinal images. We jointly train a keypoint detection and description network on small patches using a classification and a cross-modal descriptor loss function and apply the network to the full image size in the test phase. 
Our method demonstrates the best registration performance on our and a public multi-modal dataset in comparison to competing methods.
\end{abstract}

\section{Introduction}
In ophthalmological imaging, different imaging systems, such as color fundus (CF), infrared (IR), fluorescein angiography (FA), or the more recent optical coherence tomography (OCT) and OCT angiography (OCTA), are used. For the diagnosis, often multiple images that might come from different systems or capturing times are used, particularly for long-term monitoring of the progression of retinal diseases, such as diabetic retinopathy, glaucoma, or age-related macular degeneration.
Multi-modal registration techniques that accurately align the vessel structures in the different images can support ophthalmologists by allowing a direct pixel-based comparison of the images.

Multi-modal retinal registration methods estimate an affine transform, a homography, or a non-rigid displacement field.
Here, we focus on feature-based methods for homography estimation. These methods generally consist of keypoint detection, descriptor learning, and descriptor matching.
Conventional methods address the multi-modal keypoint detection and description, for instance by introducing a partial intensity invariant feature descriptor (PIIFD)~\cite{ChenJ2010} combined with the Harris corner detector. 
Deep learning methods replace all or some steps by neural networks.
In DeepSPA~\cite{LeeJ2019}, a convolutional neural network (CNN) is used to classify patches extracted at vascular junctions based on a step pattern representation.
GLAMPoints~\cite{TruongP2019} uses a U-Net as keypoint detector with root SIFT descriptors for retinal images. It is trained to maximize the keypoint matching in a self-supervised manner by homographic warping.
Wang \etal~\cite{WangY2021} proposed a weakly supervised learning-based pipeline composed of a multi-modal retinal vessel segmentation network, SuperPoint~\cite{DeToneD2018}, and an outlier network based on homography estimation. 

In this paper, we propose a deep learning method for multi-modal retinal image registration based on jointly learning a keypoint detection and description network that extracts features of the vessel structure. 
We base our approach on CraquelureNet~\cite{SindelA2021} and transfer the task of learning a cross-modal keypoint detector and descriptor based on crack structures in paintings to the medical domain. For this task, we created a multi-modal dataset with manual keypoint pair and class annotations.

\begin{figure}[t]
	\setlength{\figbreite}{0.49\textwidth}
	\centering
	\caption{Our multi-modal retinal image registration method using a CNN to extract deep features of the vessel structures.}
	\subfigure[Training]{\includegraphics[width=\figbreite]{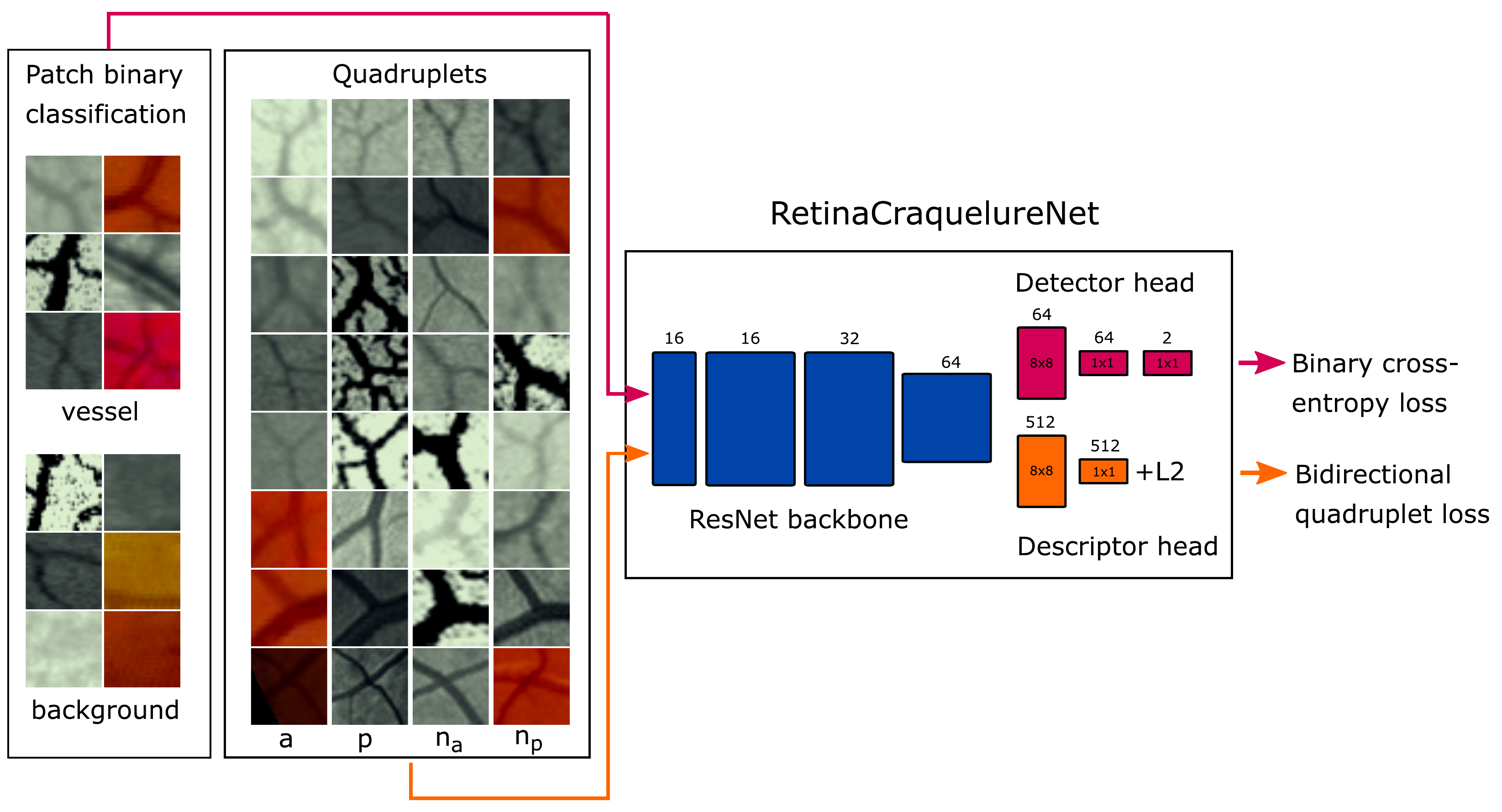}}
	\hfill
	\subfigure[Inference]{\includegraphics[width=\figbreite]{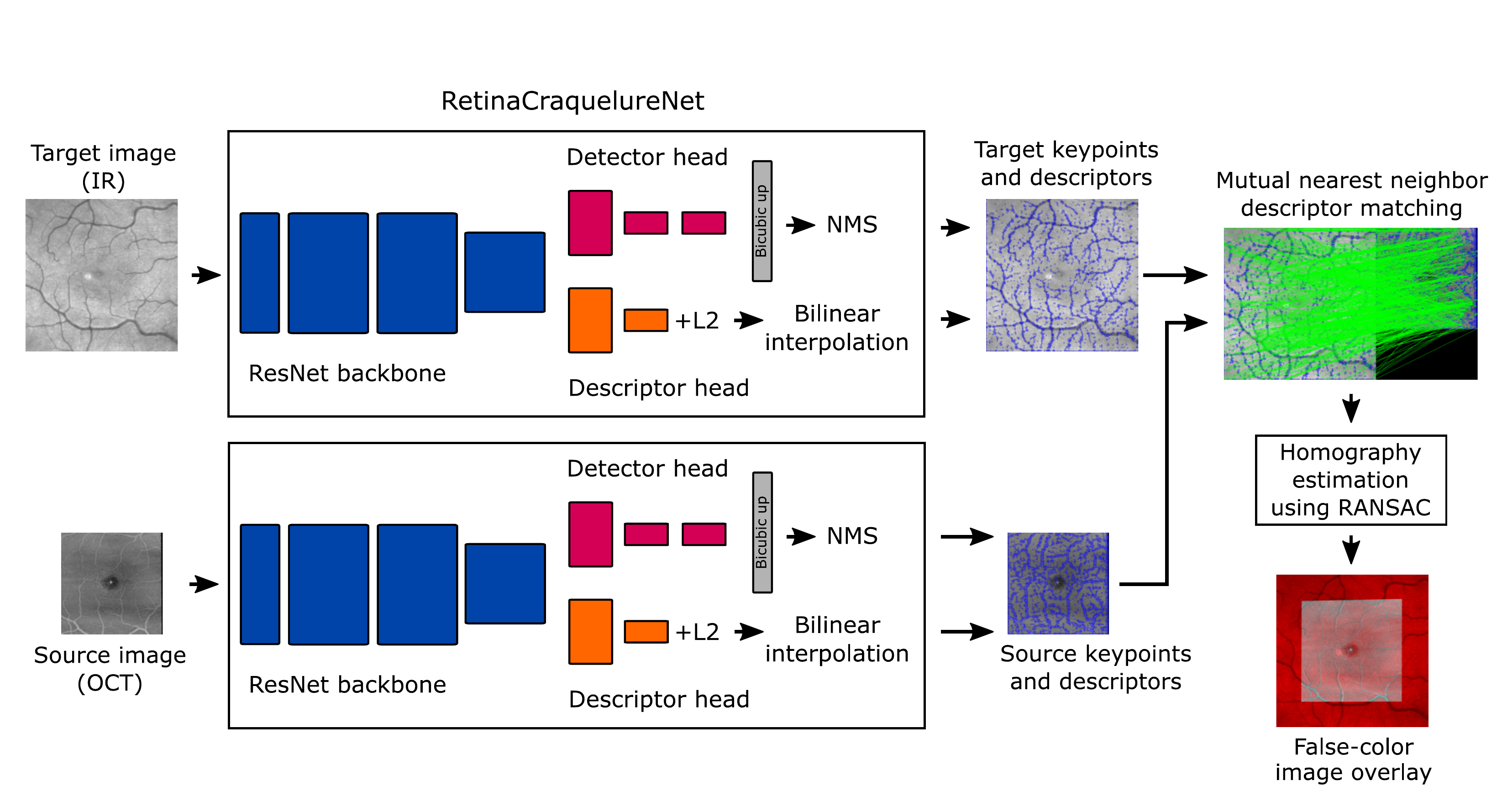}}
	\label{fig-01}
\end{figure} 

\section{Materials and methods}
\subsection{Network for multi-modal keypoint detection and description}
We adopt the CraquelureNet~\cite{SindelA2021} as network architecture for our task of multi-modal retinal image registration, named RetinaCraquelureNet, as shown in Figure~\ref{fig-01}. 
The fully-convolutional CraquelureNet consists of a ResNet~\cite{HeK2016} backbone and two heads.

\vspace{0.5em}
\noindent\textbf{Detector and descriptor loss functions:}
The keypoint detection and description heads are jointly trained with small image patches of size $32 \times 32 \times 3$ using the multi-task loss~\cite{SindelA2021}:
$\mathcal{L}_{\text{Total}} = \lambda_{\text{Det}} \mathcal{L}_{\text{BCE}} + \lambda_{\text{Desc}} \mathcal{L}_{\text{QuadB}}$,
where $\lambda_{\text{Det}},\lambda_{\text{Desc}}$ are the weights for the detector and descriptor loss.

The keypoint detection head is trained using the binary cross-entropy loss $\mathcal{L}_{\text{BCE}}$ with the two classes \elqq vessel\erqq~and \elqq background\erqq, where \elqq vessel\erqq~refers to patches centered at a striking position in the vessel structure, such as a bifurcation, intersection, or a sharp bend. Analogously to~\cite{SindelA2021}, we randomly sample the same number of patches from each class and each modality per batch.

The description head is trained using the bidirectional quadruplet loss~\cite{SindelA2021}, which applies an online in-batch hard negative mining strategy to randomly sample positive pairs (anchor $a$ and positive counterpart $p$) for one batch and to select the closest non-matching descriptors in both directions within this batch~\cite{SindelA2021}:
\begin{align}
\begin{split}
\mathcal{L}_{\text{QuadB}}(a,p,n_a,n_p) = \max [0, m + d(a,p) - d(a,n_a)] 
\\+ \max [0, m + d(p,a) - d(p,n_p)],
\end{split}
\end{align}
where $m$ is the margin, $d(x,y)$ the Euclidean distance, $n_a$ the closest negative to $a$, and $n_p$ is the closest negative to $p$.

\vspace{0.5em}
\noindent\textbf{Keypoint detection, descriptor matching, and homography estimation:}
For inference, we feed the complete image to the network at once. 
The detector output is bicubically upscaled by a factor of 4 and a dense keypoint confidence heatmap is computed by 
the difference of the two class predictions of the upscaled output.
Then, non-maximum suppression with a threshold of $4$ pixel is applied and the $N_\text{max}$ keypoints with the highest confidence values are extracted and the corresponding descriptors are linearly interpolated~\cite{SindelA2021}.
Distinctive point correspondences are determined in both images by brute force mutual nearest neighbor descriptor matching and random sample consensus (RANSAC)~\cite{FischlerMA1981} (reprojection error of $5$ pixel) is applied for homography estimation.
\begin{figure}[t]
	\setlength{\figbreite}{0.9\textwidth} 
	\centering	
	\caption{Keypoint confidence heatmaps (red to blue) and extracted keypoints of our multi-modal registration method RetinaCraquelureNet for both datasets.}	
	\includegraphics[width=\figbreite]{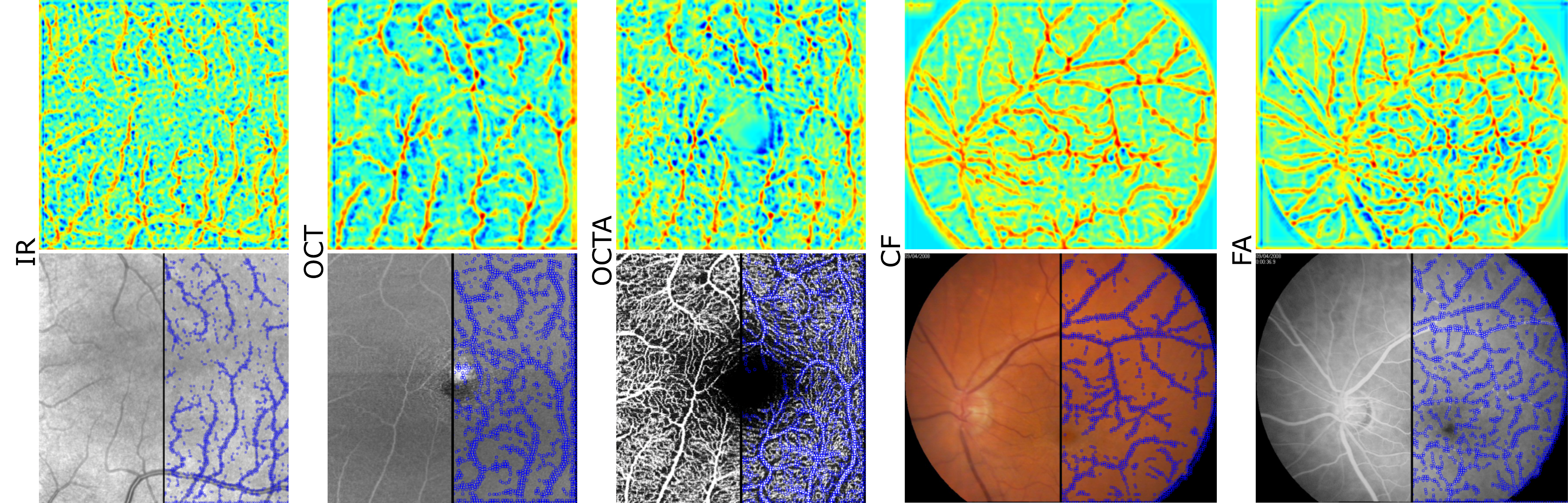}
	\label{fig-02}
\end{figure}

\subsection{Multi-modal retinal datasets}
Our IR-OCT-OCTA dataset, provided by the Department of Ophthalmology, FAU Erlangen-N\"urnberg, 
consists of multi-modal images of the macula of $46$ controls measured by Spectralis OCT II, Heidelberg Engineering.
For each control, the multi-modal images (IR images, OCT volumes, and OCTA volumes) of the same eye were scanned up to three times a day.
For this work, we use the IR image ($768 \times 768$) and the en-face OCT/OCTA projections of the SVP layer (Par off) of the macula (both $512 \times 512$). 
The image splits for each modality are: train: $89$, val: $15$, and test: $30$. Eyes of the same control are 
inside one set.

Further, we use a public dataset~\cite{ShirinH2021} of color fundus (CF, $576 \times 720\times 3$) and fluorescein angiography (FA, $576 \times 720$) images, which are composed of $29$ image pairs of controls and $30$ pairs of patients with diabetic retinopathy.
The image pair splits are: train: $35$, val: $10$, and test: $14$, where healthy and non-healthy eyes are equally distributed.

We manually annotated $N_\text{kp}$ matching keypoints at striking vessel positions in each image for each test person, where $N_\text{kp}>=21$ for IR-OCT-OCTA and $N_\text{kp}=40$ for CF-FA. 
For the keypoint detection task, we use all keypoints for the vessel class (train: 6273|2800, val: 945|800 for IR-OCT-OCTA|CF-FA) and the same number of points for the background class.
For the description task, we build the positive pairs from all multi-modal image pairs 
(train: 6273|1400, val: 945|400 for IR-OCT-OCTA|CF-FA) 
and additional for the IR-OCT-OCTA dataset from uni-modal image pairs of scans from different recording times (train: 6021, val: 945). 
For the test set, we annotated $6$ control points per image and computed ground truth homographies.

\begin{figure}[t]
	\setlength{\figbreite}{0.7\textwidth} 
	\centering	
	\caption{Qualitative results for one IR-OCTA example.}		
	\includegraphics[width=\figbreite]{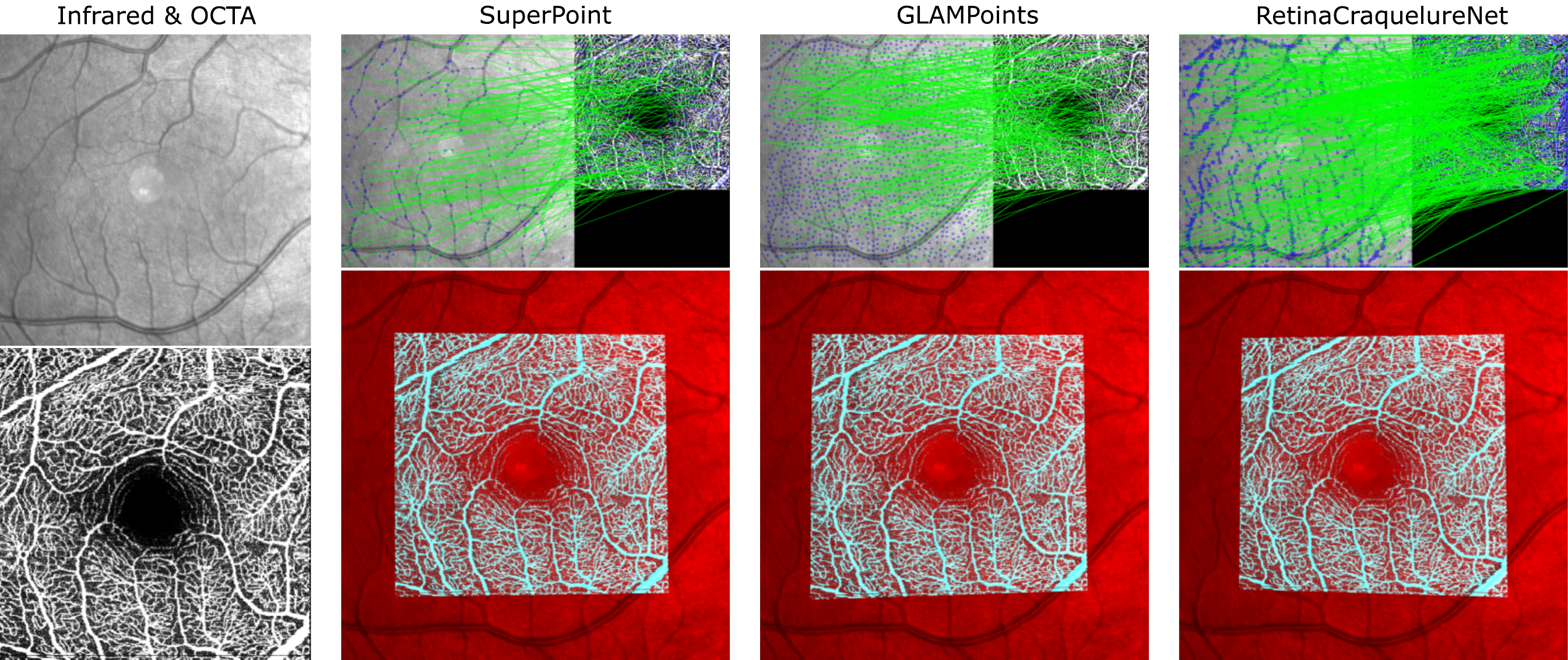}	
	\label{fig-03}
\end{figure}

\subsection{Experimental details}
We train our RetinaCraquelureNet (using pretrained weights by~\cite{SindelA2021}) for the combination of our IR-OCT-OCTA dataset and the public CF-FA dataset, where we oversample the smaller dataset.
We use Adam solver with a learning rate of $\eta=1\cdot10^{-4}$ for $20$ epochs with early stopping, a batch size of $576$ for detector and $288$ for descriptor, $m=1$, $\lambda_\text{Det}=1$, $\lambda_\text{Desc}=1$ (analogously to~\cite{SindelA2021}),
and online data augmentation (color jittering, horizontal/vertical flipping and for the keypoint pairs additional joint rotation, scaling, and cropping). 

We compare our method with SuperPoint~\cite{DeToneD2018} and GLAMPoints~\cite{TruongP2019}. We fine-tuned both methods with our combined dataset by extending the training code of~\cite{JauYY2020} for SuperPoint and~\cite{TruongP2019} for GLAMPoints 
by additionally incorporating multi-modal pairs for the homography warping and picked the best models based on the validation split.

For the experiments, we invert the images of OCT, OCTA, and FA to depict all vessels in dark.
For RetinaCraquelureNet, we feed the images in RGB, for GLAMPoints as green channel, and for SuperPoint in grayscale. 
For all methods, we use the same test settings (descriptor matching, RANSAC, $N_\text{max}=4000$).

As metrics we use the success rate of the registration based on the mean Euclidean error $\text{SR}_\text{ME} = \frac{1}{N} \sum_{i=1}^{N} (( \frac{1}{N_j} \sum_{j=1}^{N_j}  \mathrm{D}_{ij}) <= \epsilon )$ and maximum Euclidean error $\text{SR}_\text{MAE} = \frac{1}{N} \sum_{i=1}^{N} ((\max_{j \in N_j}  \mathrm{D}_{ij}) <= \epsilon )$ of the $N_j=6$ control points and the pixel error threshold $\epsilon$. Therefore, we compute the Euclidean error $\mathrm{D}_{ij} =  \| T(p_{ij},H_{\text{pred}_i}) - q_{ij} \|_2$ with $p_{ij}$ being the $j$-th source point and $q_{ij}$ the $j$-th target point, $H_{\text{pred}_i}$ the predicted homography, and $T(p_{ij},H_{\text{pred}_i})$ the projected $j$-th source point of image $i$.
Further, we compute the detector repeatability (Rep) as defined in~\cite{DeToneD2018} and the matching inlier ratio (MIR) as the fraction of number of RANSAC inliers and number of detected matches per image.

\begin{table}[t]
\caption{Quantitative evaluation for the public CF-FA dataset.} 
\label{tab-01}
\begin{scriptsize}
\begin{tabular*}{\textwidth}{l@{\extracolsep\fill}llll}
\hline
Metrics [\%] & $\text{SR}_\text{ME}$ ($\epsilon=3$) & $\text{SR}_\text{MAE}$ ($\epsilon=5$) & Rep ($\epsilon=5$) & MIR ($\epsilon=5$)\\
\hline 
SuperPoint (fine-tuned) & \textbf{100.0} & 92.9 & 53.7 & 56.5 \\ 
GLAMPoints (fine-tuned) & 78.6 & 78.6 & 35.3 & 27.6 \\
RetinaCraquelureNet & \textbf{100.0} & \textbf{100.0} & \textbf{78.4} & \textbf{69.2} \\ 
\hline 
\end{tabular*}
\end{scriptsize}
\end{table}
\begin{figure}[b]
\setlength{\figbreite}{0.27\textwidth}
\centering
\includegraphics[width=0.9\textwidth]{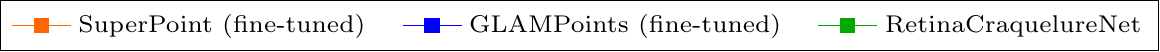} 

\subfigure[IR-OCT]{
\includegraphics[width=\figbreite]{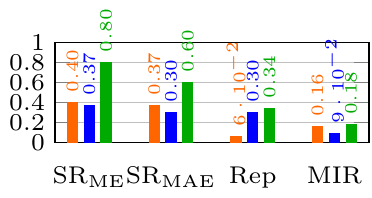}
}
\subfigure[IR-OCTA]{
\includegraphics[width=\figbreite]{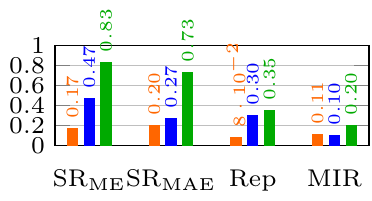}
}
\subfigure[OCT-OCTA]{
\includegraphics[width=\figbreite]{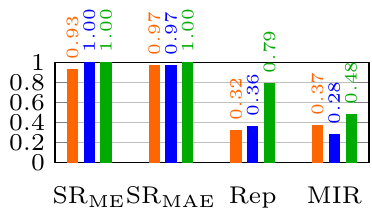}
}

\subfigure[IR-IR]{
\includegraphics[width=\figbreite]{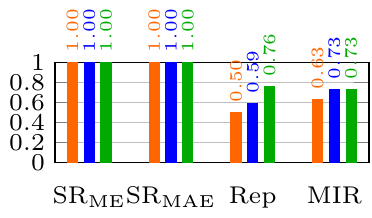}
}
\subfigure[OCT-OCT]{
\includegraphics[width=\figbreite]{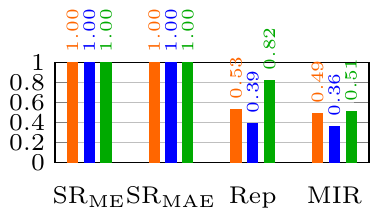}
}
\subfigure[OCTA-OCTA]{
\includegraphics[width=\figbreite]{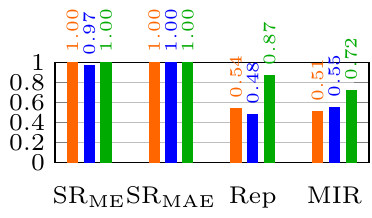}
}

\caption{Quantitative evaluation for our IR-OCT-OCTA dataset using $\text{SR}_\text{ME}$ ($\epsilon=5$), $\text{SR}_\text{MAE}$ ($\epsilon=10$), Rep ($\epsilon=5$), and MIR ($\epsilon=5$). In (c-f) we use follow-up image pairs.}
\label{fig-04}
\end{figure}

\section{Results}
Some qualitative results of our RetinaCraquelureNet are shown in Figure~\ref{fig-02} that depicts for one example of each modality the confidence keypoint heatmaps and the extracted keypoints that are concentrated on interesting points in the vessel structures. In Figure~\ref{fig-03} the registration performance of RetinaCraquelureNet, GLAMPoints, and SuperPoint is visually compared for one IR-OCTA example. RetinaCraquelureNet detects the highest number of correct matches which are densely spread over the images resulting in the most accurate image overlay. GLAMPoints detects densely distributed keypoints which are also located in the background but with fewer correct matches.
SuperPoint finds the lowest number of keypoints and correct matches. The image overlays of the competing methods show some misalignments at the left borders.

The quantitative evaluation for the IR-OCT-OCTA dataset is summarized in Figure~\ref{fig-04}. RetinaCraquelureNet clearly outperforms GLAMPoints and SuperPoint for the multi-modal image pairs in all metrics. Regarding the registration of the follow-up images, all methods reach a success rate of about $100 \, \%$ with \mbox{$\text{ME}<=5$} and $\text{MAE}<=10$, however our method obtains considerably higher scores in Rep and MIR. 
Results for the CF-FA dataset in Table~\ref{tab-01} also show the advantage of our method, which successfully registers all images with \mbox{$\text{ME}<=3$} and $\text{MAE}<=5$ and gains best scores in Rep and MIR. 

\section{Discussion}
We trained a CNN that extracts features of the vessel structure to jointly learn a cross-modal keypoint detector and descriptor for multi-modal retinal registration.
In the experiments, we showed that our method achieves the best registration results for both multi-modal datasets.
For the more challenging IR-OCT and IR-OCTA registration, which has a smaller overlap region, we still achieve good success rates while the competing methods show a strong decline.
By training our method jointly on two datasets, the network learns to detect distinctive features in five different modalities. 
Further, we demonstrated that the same trained model can be used to register multi-modal images and follow-up scans.
Thus, our method can be very beneficial to support the long time analysis of retinal disease progression. 
As future work, we will investigate deep learning methods inspired by known operators for our registration pipeline.

\bibliographystyle{bvm}

\bibliography{RetinaCraquelureNet}

\marginpar{\color{white}E\articlenumber}

\end{document}